**Shrunk loop theorem for the topology probabilities of closed Brownian (or Feynman) paths on the twice punctured plane.**

O Giraud, A Thain[†] and J H Hannay

H H Wills Physics Laboratory, University of Bristol, Tyndall Avenue, Bristol BS8 1TL, UK

[†]Present address: BAE SYSTEMS, Advanced Technology Centre – Sowerby, Filton, Bristol BS34 7QW, UK

Abstract

The shrunk loop theorem proved here is an integral identity which facilitates the calculation of the relative probability (or probability amplitude) of any given topology that a free, closed Brownian (or Feynman) path of a given 'duration' might have on the twice punctured plane (plane with two marked points). The result is expressed as a 'scattering' series of integrals of increasing dimensionality based on the maximally shrunk version of the path. Physically this applies in different contexts: (i) the topology probability of a closed ideal polymer chain on a plane with two impassable points, (ii) the trace of the Schrödinger Green function, and thence spectral information, in the presence of two Aharonov-Bohm fluxes, (iii) the same with two branch points of a Riemann surface instead of fluxes. Our theorem starts from the Stovicek scattering expansion for the Green function in the presence of two Aharonov-Bohm flux lines, which itself is based on the famous Sommerfeld one puncture point solution of 1896 (the one puncture case has much easier topology, just one winding number). Stovicek's expansion itself can supply the results at the expense of choosing a base point on the loop and then integrating it away. The shrunk loop theorem eliminates this extra two dimensional integration, distilling the topology from the geometry.

# 1. Introduction

The 'shrunk loop theorem' presented here is an integral identity conjectured by the last two authors in 2001, but not published. Numerical evaluation of the integrals for the simplest case had provided convincing evidence of the identity, but a proof was lacking. An analytical proof of the conjecture has been provided by the first author, covering all numbers of scatterings. It is involved and takes up most of this paper (section 3), but the theory itself (sections 1 and 2) is not difficult to describe as follows.

The theorem facilitates the calculation of the relative probability (or probability amplitude) of any given topology that a free, closed Brownian random walk (or Feynman path [4]) of given 'duration' might have on a twice punctured plane (plane with two marked points), conditional on the requirement that the loop encloses at least one point. It expresses the result as a 'scatter' series of integrals of the kind (2.8) based on 'shrunk' loop scatter paths. The minimal one of these loops is the path which a stretched rubber band, originally placed in the shape of the random walk, would adopt if released; further terms in the series are represented by the loops going back and forth any number of times between both points, and having the same topology (see right column of Figure 1).

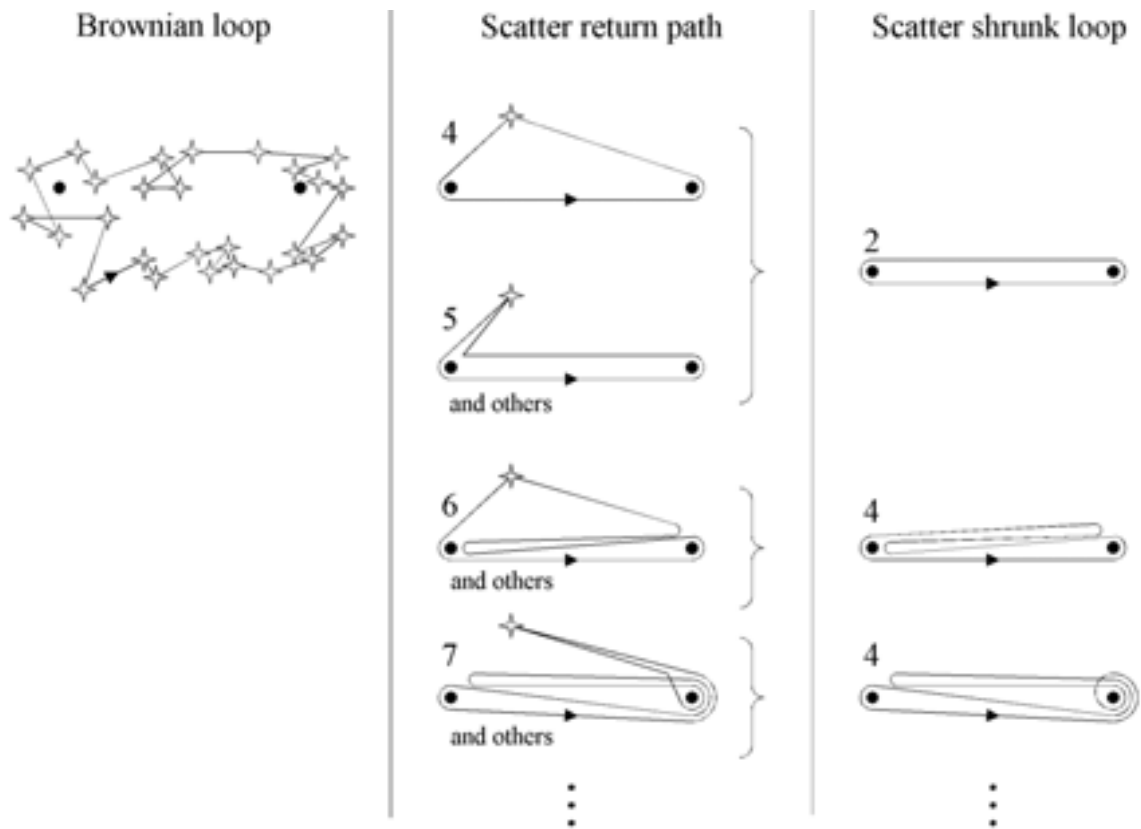

**Figure 1.** *Schematic picture of the two stages in the reduction of the (Feynman, infinite dimensional) integral over all Brownian loops of a particular topology (left). First Stovicek's scattering expansion reduces it to an infinite set of terms (middle column) each involving an integral of finite dimensionality: two spatial dimensions indicated by the star and one more for each 'scatter'. Some examples of these with their number of integrations are shown. Then (right column) our Shrunk loop theorem eliminates (i.e. performs implicitly) the two spatial integrations and combines together those pictures of the middle column which have the same shrunk shape if the star, imagined on elastic strings, is released. The intermediate legs passing back and forth between the points are supposed unchanged by the release even if they are mere u-turns. Note that the number of scatters in the pictures being combined need not be the same but can differ by one as in the top example shown.*

Physically, the theorem applies, on the one hand, directly to realizations of closed Brownian paths such as ideal polymer chain loops in a plane. On the other, with interpretation as Feynman paths (imaginary diffusion coefficient, or time) the theorem applies to wave propagation on a plane governed by the Schrödinger equation, in the presence, for example, of two Aharonov-Bohm flux lines perpendicular to the plane. Equally well it applies, by Fourier transformation, to the Helmholtz equation (the time independent wave equation) on a Riemann surface with two branch points joined by a branch cut (for instance the surface for

$\sqrt{1-z^2}$ ). In these wave cases the sum over all closed loops gives the trace of the propagator, or of the Green function, and thus spectral information. The Riemann surface result, for instance, is a first step in the quest for an exact scattering theory for the spectrum of polygon billiards.

The starting point for our analysis, in which the shrunk loop theorem only enters at the second stage, will be the scattering expression for the propagator on the twice punctured plane. The puncture points do not themselves scatter (they are just marks on the free plane, so the propagator is just the free propagator); rather, the scattering expression serves to classify the paths according to their topology. The simpler once punctured plane was solved famously by Sommerfeld in 1896 [7, 8]. This was not in the context of Brownian paths but of waves. His solution supplies the exact Green function for waves obeying the time independent wave equation, or Helmholtz equation ($\nabla^2\psi + k^2\psi = 0$) on a flattened helicoid surface (Riemann surface of log $z$). The Green function for the once punctured plane can be understood as a discrete sum over classes of paths, each having a different topology. Returning to the twice punctured plane, there is a natural generalization of the Sommerfeld solution as a scattering series due to Stovicek [9] who gave the wave (quantum) propagator and the Green function for a plane with two Aharonov-Bohm flux lines. The series can also be understood as a sum over paths, which inherits the topological character of the Sommerfeld solution.

The Green function is related to the Brownian propagator (the end-to-end displacement probability distribution) of an open Brownian path, by Fourier transformation. Specifically if the mean square end-to-end displacement is denoted by 2$t$, where $t$ is referred to as its 'time' duration, the propagator $K(\mathbf{r},\mathbf{r'},t)$ is the Fourier transform, with respect to imaginary time, of $G(\mathbf{r},\mathbf{r'},E)$ with $E=k^2/2$. The Feynman integral over all loops that we will require for the probability calculation is obtained by spatial integration of the return propagator (with coincident end points) $K(\mathbf{r},\mathbf{r},t)$. Since the presence of a special point (the coincident end point) is artificial, one should expect that performing the spatial integration over the end point position would yield a simple result of recognizable form. This is what the shrunk loop theorem achieves; the resulting expression has two fewer integrations and consists in essentially the same scattering formula applied to the shrunk loop rather than the one containing the moveable end point.

## 2. The shrunk-loop theorem

By way of introduction we first analyse the once punctured plane.

### 2.1. Once punctured plane

A flattened infinite helicoid (i.e. Riemann surface for log *z*) captures topology for the once punctured plane (it is its 'covering space'): given two layers of the helicoid, all paths going from one layer to the other have a certain topology  One circuit of the helicoid axis is distinguished from a circuit not enclosing the axis by having its end points on separate layers. The Green function $G(\mathbf{r},\mathbf{r}',E)$ for waves on a flattened infinite helicoid was famously obtained by Sommerfeld in 1896 [8]; see also Edwards [3] and Berry [1]. Following Stovicek [9], the solution for its Fourier transform, the propagator *K(r,r',t)*, can usefully be expressed in the following way, as a sum of contributions corresponding to the different topologies.

The propagator on the plane between **r** and **r'** is $K_{\text{free}}(\mathbf{r},\mathbf{r}',t) = (1/2\pi t)\exp\left(-|\mathbf{r}-\mathbf{r}'|^2/2t\right)$. This can be alternatively expressed in terms of an imagined uniform velocity **v**=(**r'**-**r**)/t as $K_{\text{free}}(\mathbf{r},\mathbf{r}',t) = (1/2\pi t)\exp(-|v|^2 t/2)$. On the helicoid, for two given points **r** and **r'**, the scattering propagator (here there genuinely is scattering on the helicoid axis) can be expressed in terms of two straight legs, one from **r** to the axis and one from the axis to **r'**, connected by a path winding around the axis. Each leg is imagined to have a hypothetical positive duration $t_j$, to be integrated over (with the condition that the sum of the durations is *t*), and thus a certain (uniform) velocity; the velocity vector will be represented by a complex number $v_j = v_{jx} + iv_{jy}$. The scattering propagator will be expressed as a product of two free propagators $(1/2\pi t_0)\exp(-|v_0|^2 t_0/2)$ and $(1/2\pi t_1)\exp(-|v_1|^2 t_1/2)$ associated with the two straight legs, and of a scattering factor $(-2\pi)/[(\ln(v_1/v_0))^2 + \pi^2]$ supplied by Sommerfeld via Stovicek, depending on the velocity ratio after and before the scatter.  The imaginary part of the logarithm is the angle between the velocity vectors and counts the windings of the path around the axis. It should be mentioned incidentally at this point perhaps, that a technical benefit of studying closed paths is that there is no ambiguity in the definition of the winding number *M*; the winding number of a closed path is convention independent so that branch cut considerations are avoided.

If the initial and final points **r** and **r'** of the path are mutually visible on the helicoid then one has to add the free space plane propagator from **r** to **r'**. Let us call *M* the number of layers of the helicoid that need to be pierced to bring **r'** onto the layer visible to **r**. The propagator for the helicoid is then equal to

$$K_M(\mathbf{r},\mathbf{r}';t) = \begin{cases} K_0(\mathbf{r},\mathbf{r}';t) + K_{\text{free}}(\mathbf{r},\mathbf{r}';t) & \text{if } \mathbf{r} \text{ and } \mathbf{r}' \text{ are on the same layer } (M=0) \quad (2.1)\\ \int_0^\infty dt_0 \int_0^\infty dt_1 \frac{1}{2\pi t_0} e^{-r_0^2/2t_0} \frac{2\pi}{\left(\phi + i\ln\frac{r_1 t_0}{r_0 t_1}\right)^2 - \pi^2} \frac{1}{2\pi t_1} e^{-r_1^2/2t_1} \delta(t_0 + t_1 - t) & \text{otherwise} \end{cases}$$

where $\phi$ is the angle between the two legs including windings ($-\infty < \phi < \infty$).

If we consider a propagation on the free infinite plane, we can still classify paths according to their winding around a fixed point in the plane. Each scattering term $K_M$ can be interpreted as a contribution to the total (non-scattering) propagator on the plane coming from the trajectories winding $M$ times around the marked point. The free propagator on the plane, which is the sum over all paths from $\mathbf{r}$ to $\mathbf{r'}$ considered on the plane, $K_{\text{free}}(\mathbf{r},\mathbf{r'},t)$, is equal to the sum over the same set of paths considered on the helicoid where the paths may arrive on any layer, that is $K_{\text{free}}(\mathbf{r},\mathbf{r'},t) + K_0(\mathbf{r},\mathbf{r'},t) + \sum_{M \neq 0} K_M(\mathbf{r},\mathbf{r'},t)$. Therefore

$$\sum_M K_M(\mathbf{r},\mathbf{r'},t) = 0. \tag{2.2}$$

Note that this property can be verified directly from Equation (2.1): the diffraction coefficient in (2.1) is of the form $1/((2M\pi + x)^2 - \pi^2)$ and gives 0 when summed over all $M$. In the case we are interested in, where $\mathbf{r}=\mathbf{r'}$ (the return propagator for a closed loop), this relation will be useful to normalise the probability for a loop to have a given topology of winding around a "puncture" (marked) point. The probability for each different topology is proportional (up to a constant) to the trace (that is, the integral over the plane) of the return propagator associated with this topology. The normalisation constant is given by the sum of contributions of all non-detached topologies $\sum_{M \neq 0} K_M(\mathbf{r},\mathbf{r},t)$ integrated over $\mathbf{r}$. According to (2.2), this sum is just minus the detached $M=0$ value, $K_0(\mathbf{r},\mathbf{r},t)$.

For the once punctured plane the probabilities $\int K_M$ are easy to evaluate analytically and we do so now in useful preparation for the twice punctured plane. For the term with winding number $M$ the trace $\int K_M$ of the propagator, using polar coordinates, is

$$\int_0^\infty dr \int_0^{2\pi} rd\theta \int_0^\infty dt_0 \int_0^\infty dt_1 \frac{1}{2\pi t_0} \exp[-\tfrac{1}{2} r^2/t_0]$$
$$\times \frac{-2\pi}{[(\ln(t_0/t_1) - i2M\pi)^2 + \pi^2]} \frac{1}{2\pi t_1} \exp[-\tfrac{1}{2}r^2/t_1]\, \delta(t_0 + t_1 - t)$$
$$= \int_0^\infty dr \int_0^{2\pi} rd\theta \int_0^\infty dt' \int_{-\infty}^\infty ds\, \frac{1}{2\pi}\frac{1}{2\pi t'} \exp[-2r^2(\cosh\tfrac{s}{2})^2/t'] \times \frac{-2\pi}{[(s - i2M\pi)^2 + \pi^2]}\, \delta(t' - t) \qquad (2.3)$$

after the change of variables $s=\ln(t_0/t_1)$, and $t'=t_0+t_1$, giving Jacobian $t_0 t_1/t'$. The $r$ and $\theta$ and then $t'$ integrals can be evaluated and the denominator can be split by partial fractions.

$$\int_{-\infty}^\infty ds\, \frac{1}{8\pi} \frac{1}{(\cosh\tfrac{s}{2})^2} \times \left\{ \frac{1}{[2M\pi + i(s + i\pi)]} - \frac{1}{[2M\pi + i(s - i\pi)]} \right\} \tag{2.4}$$

Shifting $s$ by $\pm i(\pi - 0)$ by a change of variables, oppositely for each of these terms to make the denominators equal, gives

$$\left[\int_{-\infty-i\pi}^{\infty-i\pi}ds - \int_{-\infty+i\pi}^{\infty+i\pi}ds\right]\frac{1}{8\pi}\ \frac{1}{(\sinh\frac{1}{2}s)^2}\times\frac{1}{[2M\pi+is]} \qquad (2.5)$$

The pair of contours can be closed at infinity and the only pole enclosed is the double one at the origin $s=0$ for $M\neq 0$; this pole is triple for $M=0$. Finally,

$$\int d^2\boldsymbol{r}\,K_M(\boldsymbol{r},\boldsymbol{r};t) = \begin{vmatrix} 1/(4\pi^2 M^2) & M\neq 0 \\ -1/12 & M=0 \end{vmatrix} \qquad (2.6)$$

As expected, the sum over $M$ yields zero. Therefore the $M=0$ term (which is negative in contrast to the rest) serves, with its sign reversed, as a normalisation for the $M\neq 0$ contributions.

## 2.2. Twice punctured plane

In the case of the twice punctured plane, the evaluation of the relative probability for each different topology of path requires the calculation of the analogue of $K_M$, that is the contribution from each topology to the propagator $K(\mathbf{r},\mathbf{r'},t)$ of the twice punctured plane. This propagator cannot, as far as is known, be expressed in finite terms, but it was obtained as an exact scattering series by Stovicek in 1989 [9]. Actually the closely related problem in optics of wave diffraction by a slit had been solved in a different way as an exact scattering series by Schwarzchild (of relativity fame) in 1902 [6], and the results of the two series specialized to this case are equal, term by term [5]. In the Stovicek series the wave from the source point $\mathbf{r}$ scatters alternately from the two points and after some number of such scatters goes to the observation point $\mathbf{r'}$.

A contribution to the propagator $K(\mathbf{r},\mathbf{r'},t)$ corresponding to a given topology can be depicted as a sequence of straight legs, with all but the first and last being back and forth between the puncture points. As before, between successive legs the path winds around one of the scattering points some number of times. For closed paths, which will be our interest, this number is convention independent, so branch cut considerations are again unnecessary. The propagator for a given scatter sequence is formed as in the one puncture plane by ascribing a hypothetical duration to each leg and integrating over all the durations. The integrand is the product of alternating leg, scatter, leg, scatter…,leg factors described above:

$$K(\mathbf{r},\mathbf{r}';t) = \int_0^\infty dt_0 dt_1 ... dt_n \frac{1}{2\pi t_0} e^{-r_0^2/2t_0} \frac{2\pi}{\left(\phi_1 + i\ln\frac{r_1 t_0}{r_0 t_1}\right)^2 - \pi^2} \frac{1}{2\pi t_1} e^{-r_1^2/2t_1} \times$$

$$\times \frac{2\pi}{\left(\phi_n + i\ln\frac{r_n t_{n-1}}{r_{n-1} t_n}\right)^2 - \pi^2} \frac{1}{2\pi t_n} e^{-r_n^2/2t_n} \delta(t_0 + t_1 + ... + t_n - t) \quad (2.7)$$

Here $\phi_i$ are the angles between consecutive legs, including windings ($-\infty < \phi_i < \infty$), and $r_j$ are the lengths of the successive legs. After setting the ln of the ratios of speeds equal to new variables $s_j$, the propagator reads

$$\frac{1}{2\pi t} \int_{-\infty}^\infty ds_1 ds_2 ... ds_n e^{-\frac{R^2}{2t}} \prod_{i=1}^n \frac{1}{(\phi_i + is_i)^2 - \pi^2}, \quad (2.8)$$

where $R^2 = R^2(s_1, s_2, ..., s_n)$ is defined by

$$R^2 \equiv \left(r_0 + r_1 e^{s_1} + r_2 e^{s_1+s_2} + ... + r_n e^{s_1+s_2+...+s_n}\right)\left(r_0 + r_1 e^{-s_1} + r_2 e^{-s_1-s_2} + ... + r_n e^{-s_1-s_2-...-s_n}\right). \quad (2.9)$$

and $r_1 = r_2 = ... = r_{n-1}$ is the distance between the two points. Again, the propagator on the covering space "captures" the topology, and the trace of the return propagator can be seen as the sum over all the paths going from $\mathbf{r}$ to itself on the twice punctured plane with a given number of windings around the two points.

### 2.3. Shrunk-loop theorem

We are interested in the probability for a closed path to have a given topology. Each probability is a sum of contributions which can be written as the trace of return propagator $\int K(\mathbf{r},\mathbf{r};t)$, where $K$ is of the form of Equation (2.8). The trace are of two types, as illustrated in Figure 2, depending on whether the number of scatters is even or odd. For the even type the trace term has its first and last legs from different scatter points, whereas for the odd type the first and last legs come from the same scatter point. To obtain the probability, we sum over all terms that give the same shape when their first and last leg is shrunk (see Figure 2). The shrunk-loop theorem, which will be proved in Section 3, states that the traces of terms of the form (2.8) add up to terms of the form (2.10), which have a natural diagrammatic interpretation as shrunk scatter loops. The probability for a closed path to have a given topology will then be a sum over all the "shrunk" terms (2.10) corresponding to the same topology (see the third column in Figure 1).

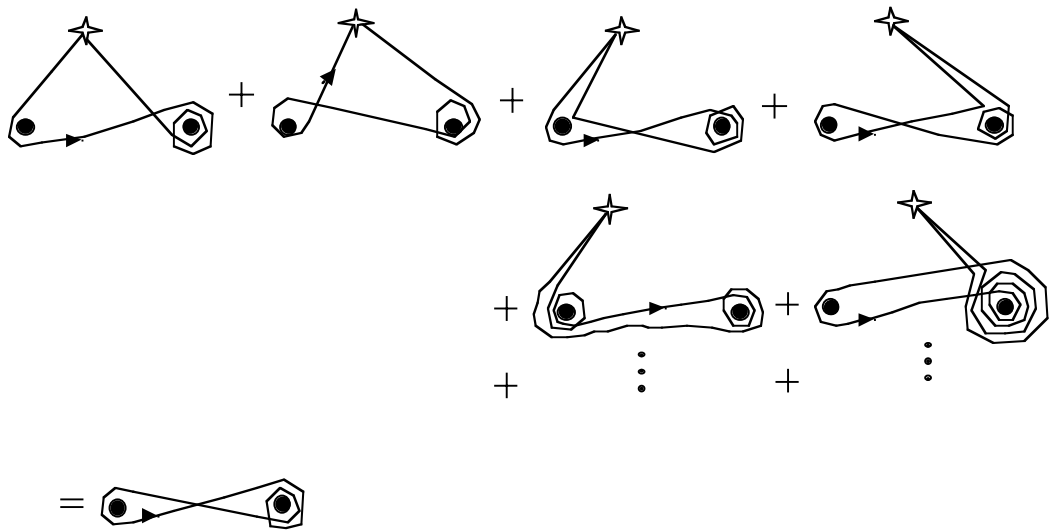

**Figure 2.** *The shrunk loop theorem is shown here for a low scatter case with winding sequence [1,-2]. An expression associated with the shrunk scatter loop path on the right has been proved equal to the sum of the contributions of all the 'related' trace integrations over the plane symbolized by stars. 'Related' means having the same shrunk shape if the star is imagined released with the first and last legs, attached to it considered as stretched elastic strings. The first two terms on the left are even type (2 scatters). The rest are odd type (3 scatter) terms. There is not just one, but an infinite sequence of odd terms associated with each of the two scatter points, with the double string from the star winding any number of times around the scattering point.*

Before stating the theorem, we need to enumerate the terms that will give the same shape when their first and last leg is shrunk. Let $2a$ be the distance between the two puncture points. Consider a scatter shrunk loop. Define $[M_1 M_2 ... M_{2n}]$ as being its successive anticlockwise winding numbers (eg those of the right hand column of Figure 1 are [1,1]; [1,1,0,0] or [0,0,1,1]; [1,2,0,-1] or [0,-1,1,2]). We will call this sequence the "scatter winding sequence".

There are as many even scatter terms as there are legs of the shrunk scatter loop (two in the case of Figure 2), since the star (representing the starting and end-point) can be associated with any one of the legs : this corresponds to different cyclic permutations of the scatter winding sequence. For the odd scatter terms, once again there is an association between the star and any one of the scatters, the star interrupting, as it were, a winding. The choice of the different scatters corresponds to cyclic permutations of the scatter winding sequence. Actually each such interruption can happen in an infinite number of ways as shown in the figure, and the shrunk loop theorem sums their contributions.

The theorem can now be stated: if $S_{[M_1M_2...M_{2n}]}$ is the sum over cyclic permutations of the indices of the even trace terms $E_{12...2n}$ (whose analytical expression is given by Equation (3.41)) and of the odd trace terms $O_{12...2n}$ (given by Equation (3.42)), then

$$S_{[M_1M_2...M_{2n}]} = \int_{-\infty}^{\infty} ds_1 ds_2 ... ds_{2n} \exp\left[-\frac{1}{2t}\left(2a + 2ae^{s_2} + ... + 2ae^{s_2+...+s_{2n}}\right)\left(2a + 2ae^{-s_2} + ... + 2ae^{-s_2-...-s_{2n}}\right)\right]$$
$$\times \prod_{i=1}^{2n} \frac{1}{(2M_i\pi + is_i)^2 - \pi^2} \times \delta(s_1 + s_2 + ... + s_{2n}). \qquad (2.10)$$

and it turns out that this expression can be read directly from the picture of the scatter shrunk loop: it is exactly the scatter propagator (2.8) for **r=r'** with its integrand multiplied by $2\pi t\delta$.

It should be remarked that in the 'semiclassical' limit (which is the limit of short duration *t*, or equivalently, widely separated puncture points) the shrunk loop theorem reduces to the formulas of semiclassical mechanics [2,5,10] for the trace of the Green function. There is then fast convergence of the scattering series with the higher terms vanishing rapidly.

### 2.4. Probabilities and normalisation

As we have already stated, a task remains after the application of the shrunk loop theorem, and separate from it, in order to find the probability of a topology, namely the enumeration of the different shrunk loop diagrams (each, by use of the theorem, having summed a collection of scatter diagrams) corresponding to the topology, as in Figure 1. The shrunk Brownian loop itself gives the minimal shrunk scatter loop (since the same shape arises from shrinking the two end legs of the closed path associated with the minimal return scatter propagator with the specified topology). It has a winding sequence comprising an even number of winding integers with no zeros, and with even cyclic permutations counted as equivalent since any (leftward) leg could be considered the starting leg. Higher terms are associated likewise with shrinking the two end legs of higher scatter paths. These have the same topology but contain u-turns, that is, turns with zero windings (bottom right picture of Figure 1, for instance). Therefore their scatter winding sequence comprises an even number of winding integers with zeros allowed (again with even cyclic permutations counted as equivalent). Each zero corresponds to a u-turn.

The topology winding sequence for any given scatter winding sequence is found by applying a straightforward reduction rule to remove the zeros (i.e. u-turns): any zero and its two neighbour integers can be replaced by a single integer, their sum. For example, [1,2,0,-1] gives the topology [1,1]. The reduction rule may need applying more than once, thus a shrunk loop winding sequence [3,0,1,-8,0,0] has six legs and reduces to [4,-8,0,0] and then to [4,-8] which specifies the topology (the order in which zeros are eliminated does not matter). What we require to generate all shrunk loop scatter sequences from the minimal one representing the topology is the reverse application of this rule, expanding a winding number $M$ to a winding sequence $N,0,M-N$ and applying this expansion repeatedly. An explicit algorithm to generate each different sequence only once (with even cyclic permutations equivalent) would be desirable, but in favourable circumstances (semiclassical limit $a^2/t \gg 1$) the contribution of shrunk scatter loops with many extra legs forming u-turns (that is with many zeros) is small and can be neglected. Moreover for the terms obtained by a single application of the expansion rule, one extra back and forth, the sum over $N$ can be evaluated by use of the result $\sum_m 1/(a+m) = \pi \cot \pi a$. It remains to find the normalization, if absolute rather than just relative probabilities are required.

The normalisation is found by the same trick as was used in the once punctured plane. There the sum of contributions of *all* scatter winding numbers $M$ with $-\infty < M < \infty$ (that is, the scatter part of the propagator on the plane) was zero, and therefore the contribution of the detached shrunk scatter loop with winding number zero equalled (minus) the sum of all the rest. Similarly now on the twice punctured plane, the propagator must have zero scatter part (because the punctures are just marks). That is, the detached topology scatter contributions must equal minus the sum of the attached ones. After spatial integration, therefore, it acts as a normalisation for the absolute probabilities of the attached loops. The detached topology scatters begin with the non-winding single scatters. These contributions are evaluated separately for each of the two points, ignorant of the other, just as for the once punctured plane above. They give the contribution $2 \times 1/3$. To this must be added the sum of contributions for two or more scatters. These come from all detached shapes of shrunk scatter loops ([0,0]; [0,0,0,0], [0,$m$,0,-$m$] and [$m$,0,-$m$,0] for $m \neq 0$; and shapes of more than four legs). So the normalization evaluation is no more difficult than the evaluation of any other individual topology. Again, in favourable cases, only the lowest-order scattering sequences will contribute significantly to the sum.

## 3. Proof of the theorem

We will now prove the shrunk loop theorem for any given scattering winding sequence $[M_1 M_2 ... M_{2n}]$, beginning with the slightly simpler case where the winding sequence is given by only two integers. An illustration of this two-scatter case is given in Figure 2.

### 3.1. Two-scatter case.

We consider paths that are winding $M_1$ times around the left obstruction point and $M_2$ times around the right one. The shrunk-loop theorem states that $E + O = S$, where all these are functions of the continuous variables $t$ and $a$ with the following definitions:

Even term $E = E_{12} + E_{21}$, with

$$E_{12} = \int_{-\infty}^{\infty} \frac{dxdy}{2\pi t} \int_{-\infty}^{\infty} ds_1 ds_2 \exp\left[-\frac{1}{2t}(r_0 + r_1 e^{s_1} + r_2 e^{s_1+s_2})(r_0 + r_1 e^{-s_1} + r_2 e^{-s_1-s_2})\right] \qquad (3.1)$$

$$\times \frac{1}{(2M_1\pi + i(s_1 + i\theta_1))^2 - \pi^2} \frac{1}{(2M_2\pi + i(s_2 + i\theta_2))^2 - \pi^2},$$

with the meaning of the symbols supplied by the figure,

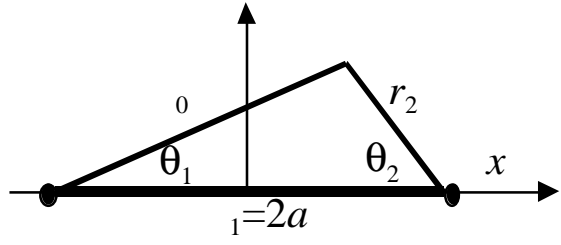

Odd term $O = O_{12} + O_{21}$, with

$$O_{12} = \int_0^{\infty} \int_0^{2\pi} \frac{rdrd\theta}{2\pi t} \int_{-\infty}^{\infty} ds_1 ds_2 ds_3 \times \qquad (3.2)$$

$$\exp\left[-\frac{1}{2t}(r_0 + r_1 e^{s_1} + r_2 e^{s_1+s_2} + r_3 e^{s_1+s_2+s_3})(r_0 + r_1 e^{-s_1} + r_2 e^{-s_1-s_2} + r_3 e^{-s_1-s_2-s_3})\right]$$

$$\times \sum_{N=-\infty}^{\infty} \frac{1}{(2N\pi + i(s_1+i\theta))^2 - \pi^2} \frac{1}{(2M_2\pi + is_2)^2 - \pi^2} \frac{1}{(2(M_1-N)\pi + i(s_3-i\theta))^2 - \pi^2}$$

with the meaning of the symbols supplied by the figure,

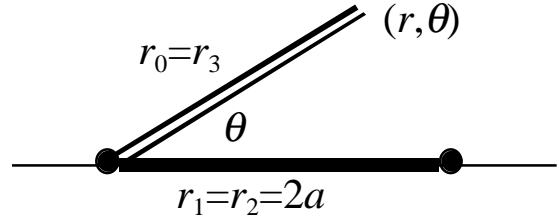

and shrunk-loop term

$$S = \int_{-\infty}^{\infty} ds_1 ds_2 \exp\left[-\frac{1}{2t}(r_1 + r_2 e^{s_2})(r_1 + r_2 e^{-s_2})\right] \times$$

$$\times \frac{1}{(2M_1\pi + is_1)^2 - \pi^2} \frac{1}{(2M_2\pi + is_2)^2 - \pi^2} \delta(s_1 + s_2) \qquad (3.3)$$

with the meaning of the symbols supplied by the figure,

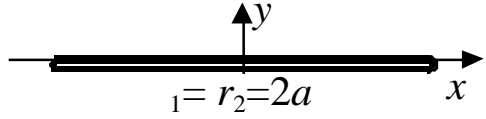

To prove the desired equality both O and E will be reduced to double integrals, like S.

*a. Reduction of the even-term E:*

The strategy here is to remove the position dependence (i.e. $x,y$ or $\theta_1,\theta_2$) from the denominator terms of the integrand and put it instead into the exponent and the limits of integration. By suitable shifting, the dependence of the limits on position can be reduced first to a discrete function of position and then none at all. The integral over the $x,y$ plane can then be evaluated leaving only the integrals over $s_1$ and $s_2$.

Introduce, following equation (2.7),

$$R^2(s_1,s_2) \equiv (r_0 + r_1 e^{s_1} + r_2 e^{s_1+s_2})(r_0 + r_1 e^{-s_1} + r_2 e^{-s_1-s_2}). \tag{3.4}$$

Then for any $\phi_1, \phi_2$,

$$\operatorname{Re}\left[R^2(s_1+i\phi_1, s_2+i\phi_2)\right] =$$
$$r_0^2 + r_1^2 + r_2^2 + 2r_0 r_1 \cosh s_1 \cos\phi_1 + 2r_1 r_2 \cosh s_2 \cos\phi_2 + 2r_0 r_2 \cosh(s_1+s_2)\cos(\phi_1+\phi_2). \tag{3.5}$$

If the three quantities $\phi_1$, $\phi_2$, and $\phi_1+\phi_2$ all lie between $-\pi/2$ and $\pi/2$ then this expression is positive and it is possible to move the contour of integration from $\int_{-\infty}^{\infty} ds_1 \int_{-\infty}^{\infty} ds_2$ to $\int_{-\infty+i\phi_1}^{\infty+i\phi_1} ds_1 \int_{-\infty+i\phi_2}^{\infty+i\phi_2} ds_2$.

For $x>a$, $y>0$ one has $0<\theta_1<\pi/2$ and $\pi/2<\theta_2<\pi-\theta_1<\pi$ (note that $\theta_2$ is oriented clockwise whereas $\theta_1$ is oriented the usual way) so one can take $\pi/2<\theta_2<\pi-\theta_1<\pi$ and $\pi/2<\theta_2<\pi-\theta_1<\pi$ and satisfy the constraints above.

For $-a<x<a$, $y>0$ one has $0<\theta_1<\pi/2$ and $0<\theta_2<\pi/2$ so for any $\alpha$ with $0<\alpha<\pi/2$ one can take $\phi_1 = -\theta_1 + \alpha$ and $\phi_2 = -\theta_2 + \pi/2 - \alpha$ and satisfy the constraints above. (Later $\alpha$ will be specified as 0 or $\pi/2$ for different pieces of the formula).

For $x<-a$, $y>0$ one has $\pi/2<\theta_1<\pi-\theta_2<\pi$ and $0<\theta_2<\pi/2$ and so one can take $\phi_1 = -\theta_1$ and $\phi_2 = -\theta_2 + \pi/2$ and satisfy the constraints above.

Define

$$g_M(s) \equiv \frac{1}{(2M\pi + is)^2 - \pi^2} \tag{3.6}$$

Then, moving the contours of integration with the shifts $\phi_i$ just stated, followed by the change of variables $s_i + i\theta_i \to s_i$ yields a set of formulas with the only position dependence in the exponent.

For $x>a$, $y>0$,

$$E_{12}^{(x>a,y>0)} = \frac{1}{2\pi t}\int_a^\infty dx \int_0^\infty dy \int_{-\infty}^\infty ds_1 \int_{-\infty+i\pi/2}^{\infty+i\pi/2} ds_2 \, \exp[-\tfrac{1}{2t}R^2(s_1-i\theta_1, s_2-i\theta_2)]\, g_{M_1}(s_1)\, g_{M_2}(s_2) \tag{3.7}$$

For $-a<x<a$, $y>0$,

$$E_{12}^{(-a<x<a,y>0)} = \frac{1}{2\pi t}\int_{-a}^a dx \int_0^\infty dy \int_{-\infty+i\alpha}^{\infty+i\alpha} ds_1 \int_{-\infty-i\alpha+i\pi/2}^{\infty-i\alpha+i\pi/2} ds_2 \, \exp[-\tfrac{1}{2t}R^2(s_1-i\theta_1, s_2-i\theta_2)]\, g_{M_1}(s_1)\, g_{M_2}(s_2) \tag{3.8}$$

For $x<-a$, $y>0$,

$$E_{12}^{(x<-a,y>0)} = \frac{1}{2\pi t}\int_{-\infty}^{-a} dx \int_0^\infty dy \int_{-\infty+i\pi/2}^{\infty+i\pi/2} ds_1 \int_{-\infty}^\infty ds_2 \, \exp[-\tfrac{1}{2t}R^2(s_1-i\theta_1, s_2-i\theta_2)]\, g_{M_1}(s_1)\, g_{M_2}(s_2). \tag{3.9}$$

Now using the fact that $r_0 \exp i\theta_1 = x+a+iy$ and $r_2 \exp i(\pi-\theta_2) = x-a+iy$,

$$R^2(s_1-i\theta_1, s_2-i\theta_2)]$$
$$\equiv (r_0 + r_1 e^{s_1-i\theta_1} + r_2 e^{s_1-i\theta_1+s_2-i\theta_2})(r_0 + r_1 e^{-s_1+i\theta_1} + r_2 e^{-s_1+i\theta_1-s_2+i\theta_2})$$
$$= -4\left[(x\sinh\tfrac{s_1+s_2}{2} - a\sinh\tfrac{s_1-s_2}{2})^2 + (y\sinh\tfrac{s_1+s_2}{2} + 2ia\cosh\tfrac{s_1}{2}\cosh\tfrac{s_2}{2})^2\right] \tag{3.10}$$

so the in all cases the $x$ and $y$ integrals separate and the $y$ one, having the same limits for all cases, can usefully be evaluated. Define

$$\chi(s_1,s_2) \equiv \frac{1}{2\pi t}\int_0^\infty dy \exp[\tfrac{2}{t}(y\sinh\tfrac{s_1+s_2}{2} + 2ia\cosh\tfrac{s_1}{2}\cosh\tfrac{s_2}{2})^2]\, g_{M_1}(s_1)g_{M_2}(s_2)$$
$$= (i\sqrt{\tfrac{t}{2}})\frac{1}{\sinh\tfrac{s_1+s_2}{2}}\tfrac{\sqrt{\pi}}{2}\mathrm{Erfc}(\lambda \cosh\tfrac{s_1}{2}\cosh\tfrac{s_2}{2})\, g_{M_1}(s_1)g_{M_2}(s_2) \tag{3.11}$$

where $\lambda \equiv 2a\sqrt{2/t}$. [Note that convergence is assured because $\mathrm{Im}\tfrac{1}{2}(s_1+s_2)=\pi/4$ and for any real $u$, $\sinh(u+i\pi/4) = -i(-\cosh u + i\sinh u)/\sqrt{2}$, which lies within an angle $\pi/4$ of the positive imaginary axis, so $\sinh^2(\tfrac{1}{2}(s_1+s_2))$ has a negative real part]. Also the $x$ integrals can be standardised by setting $i\sqrt{2/t}(x\sinh[(s_1+s_2)/2] - a\sinh[(s_1-s_2)/2])$ and inserting appropriate limits in the equation

$$\int dx \exp\left[\tfrac{2}{t}(x\sinh\tfrac{s_1+s_2}{2} - a\sinh\tfrac{s_1-s_2}{2})^2\right] = (-i\sqrt{\tfrac{t}{2}})\frac{1}{\sinh\tfrac{s_1+s_2}{2}}\int dx' \exp(-x'^2). \tag{3.12}$$

The three contributions may be thus be written as follows:

For $x>a$, $y>0$,

$$E_{12}^{(x>a,y>0)} = \int_{-\infty}^{\infty} ds_1 \int_{-\infty+i\pi/2}^{\infty+i\pi/2} ds_2\, \chi(s_1,s_2)(-i\sqrt{\tfrac{t}{2}}) \frac{1}{\sinh\frac{s_1+s_2}{2}} \left( \int_{i\lambda\cosh(s_1/2)\sinh(s_2/2)}^{i\infty \sinh\frac{s_1+s_2}{2}} dx'\exp(-x'^2) \right). \quad (3.13)$$

For $-a<x<a$, $y>0$,

$$E_{12}^{(-a<x<a,y>0)} = \int_{-\infty+i\alpha}^{\infty+i\alpha} ds_1 \int_{-\infty+i(\pi/2-\alpha)}^{\infty+i(\pi/2-\alpha)} ds_2\, \chi(s_1,s_2)(-i\sqrt{\tfrac{t}{2}}) \frac{1}{\sinh\frac{s_1+s_2}{2}} \left( \int_{-i\lambda\sinh(s_1/2)\cosh(s_2/2)}^{i\lambda\cosh(s_1/2)\sinh(s_2/2)} dx'\exp(-x'^2) \right)$$

$$= \int_{-\infty}^{\infty} ds_1 \int_{-\infty+i\pi/2}^{\infty+i\pi/2} ds_2\, \chi(s_1,s_2)(-i\sqrt{\tfrac{t}{2}}) \frac{1}{\sinh\frac{s_1+s_2}{2}} \left( \int_{0}^{i\lambda\cosh(s_1/2)\sinh(s_2/2)} dx'\exp(-x'^2) \right) \quad (3.14)$$

$$+ \int_{-\infty+i\pi/2}^{\infty+i\pi/2} ds_1 \int_{-\infty}^{\infty} ds_2\, \chi(s_1,s_2)(-i\sqrt{\tfrac{t}{2}}) \frac{1}{\sinh\frac{s_1+s_2}{2}} \left( \int_{0}^{-i\lambda\sinh(s_1/2)\cosh(s_2/2)} dx'\exp(-x'^2) \right).$$

For $x<-a$, $y>0$,

$$E_{12}^{(x<-a,y>0)} = \int_{-\infty+i\pi/2}^{\infty+i\pi/2} ds_1 \int_{-\infty}^{\infty} ds_2\, \chi(s_1,s_2)(-i\sqrt{\tfrac{t}{2}}) \frac{1}{\sinh\frac{s_1+s_2}{2}} \left( \int_{-i\infty \sinh\frac{s_1+s_2}{2}}^{-i\lambda\cosh(s_1/2)\sinh(s_2/2)} dx'\exp(-x'^2) \right). \quad (3.15)$$

All these combine to give

$$E_{12}^{(y>0)} = \left( \int_{-\infty}^{\infty} ds_1 \int_{-\infty+i\pi/2}^{\infty+i\pi/2} ds_2 + \int_{-\infty+i\pi/2}^{\infty+i\pi/2} ds_1 \int_{-\infty}^{\infty} ds_2 \right) \chi(s_1,s_2)(-i\sqrt{\tfrac{t}{2}}) \frac{1}{\sinh\frac{s_1+s_2}{2}} \left( \int_{0}^{i\infty \sinh\frac{s_1+s_2}{2}} dx'\exp(-x'^2) \right). \quad (3.16)$$

Now the limit $i\infty\sinh((s_1+s_2)/2)$ can be taken as $-\infty$ because $\mathrm{Im}(s_1+s_2)/2 = \pi/4$ and $i\sinh(u+i\pi/4) = (-\cosh u + i\sinh u)/\sqrt{2}$ which lies within an angle $\pi/4$ of the negative real axis. Thus the $x'$ integral yields $-\sqrt{\pi}/2$. Inserting the definition of $\chi$, finally, since $E = E_{12} + E_{21}$,

$$E^{(y>0)} = -\frac{1}{8} \left( \int_{-\infty}^{\infty} ds_1 \int_{-\infty+i\pi/2}^{\infty+i\pi/2} ds_2 + \int_{-\infty+i\pi/2}^{\infty+i\pi/2} ds_1 \int_{-\infty}^{\infty} ds_2 \right) \frac{g_{M_1}(s_1)\, g_{M_2}(s_2)}{\sinh^2(\frac{s_1+s_2}{2})} \mathrm{Erfc}(\lambda \cosh\tfrac{s_1}{2} \cosh\tfrac{s_2}{2}) \quad (3.17).$$

For $y<0$ the symbol $i\pi/2$ in the limits is replaced by $-i\pi/2$.

*b. Reduction of the odd-term O:*

The first step here is to perform the following change of variables:

$$\begin{cases} s_1 + s_3 \to s_1 \\ s_1 - s_3 \to s_3 \end{cases} \tag{3.18}$$

This change of variables will prove convenient later on, and it has the physical interpretation that the complex angle $s_1$ associated with the first scatter in the solution term is the sum of the complex angles $s_1$ and $s_3$ (in Equation (3.2), i.e. before the changes of variables) associated with the first and the last scatters in the odd-term.

The strategy then is to perform the integration over $\theta$ Only the diffraction coefficient depends on $\theta$; its integration gives

$$\int_0^{2\pi} d\theta \sum_{N=-\infty}^{\infty} \frac{1}{\left(-\theta + 2N\pi + i\frac{s_1+s_3}{2}\right)^2 - \pi^2} g_{M_2}(s_2) \frac{1}{\left(\theta + 2(M_1-N)\pi + i\frac{s_1-s_3}{2}\right)^2 - \pi^2}$$

$$= g_{M_2}(s_2) \int_{-\infty}^{\infty} d\theta \frac{1}{\left(-\theta + i\frac{s_1+s_3}{2}\right)^2 - \pi^2} \frac{1}{\left(\theta + 2M_1\pi + i\frac{s_1-s_3}{2}\right)^2 - \pi^2}. \tag{3.19}$$

There are four poles in the right-hand-side integrand, at $\pm\pi + i(s_1+s_3)/2$ and $(2M_1 \pm 1)\pi + i(s_1-s_3)/2$. Closing the contour of the integral over $\theta$ at infinity, we can see that the value of the integral is 0 if $s_1^2 < s_3^2$, and

$$i\,\text{sgn}(s_1)\, g_{M_2}(s_2)\left(\frac{1}{2M_1\pi + is_1}\frac{1}{2(M_1+1)\pi + is_1} - \frac{1}{2(M_1-1)\pi + is_1}\frac{1}{2M_1\pi + is_1}\right) \tag{3.20}$$

otherwise. Taking into account the Jacobien 1/2 yielded by the change of variables (3.18), the odd-term is equal to

$$O_{12} = \frac{1}{4\pi t} \int_0^\infty dr r \int_{-\infty}^\infty ds_1 \int_{-\infty}^\infty ds_2 \int_{-s_1}^{s_1} ds_3 \exp\left[-\frac{1}{2t}R^2\left(\frac{s_1+s_3}{2}, s_2, \frac{s_1-s_3}{2}\right)\right] g_{M_2}(s_2) h_{M_1}(s_1), \tag{3.21}$$

where we have introduced

$$h_{M_1}(s_1) \equiv i\left(\frac{1}{2M_1\pi + is_1}\frac{1}{2(M_1+1)\pi + is_1} - \frac{1}{2(M_1-1)\pi + is_1}\frac{1}{2M_1\pi + is_1}\right)$$
$$= i\left(g_{M_1}(s_1 - i\pi) - g_{M_1}(s_1 + i\pi)\right). \tag{3.22}$$

The integral over $s_3$ and $r$ in (3.21) can be reduced: since $r_0 = r_3 = r$ and $r_1 = r_2 = 2a$,

$$R^2\left(\frac{s_1 + s_3}{2}, s_2, \frac{s_1 - s_3}{2}\right) = 16a^2\cosh^2\frac{s_2}{2} + 16ar\cosh\frac{s_2}{2}\cosh\frac{s_1+s_2}{2}\cosh\frac{s_3}{2} + 4r^2\cosh^2\frac{s_1+s_2}{2}, \tag{3.23}$$

and Appendix A1 then yields

$$\int_{-s_2}^{s_2} ds_3 \int_0^\infty dr\, r \exp\left[-\frac{1}{2t}R^2\left(\frac{s_1+s_3}{2}, s_2, \frac{s_1-s_3}{2}\right)\right] = \frac{t}{\cosh^2\frac{s_1+s_2}{2}}\Phi(s_1, s_2), \tag{3.24}$$

where as previously we have set $\lambda = 2a\sqrt{2/t}$, and

$$\Phi(s_1, s_2) \equiv \int_1^\infty dz\, e^{-\lambda^2\left(\cosh^2\frac{s_2}{2}\right)z^2} \frac{\sinh\frac{s_1}{2}}{z\sqrt{z^2 + \sinh^2\frac{s_1}{2}}}. \tag{3.25}$$

The odd-term is therefore equal to

$$O_{12} = \frac{i}{4\pi}\int_{-\infty}^\infty ds_1 \int_{-\infty}^\infty ds_2 \frac{1}{\cosh^2\frac{s_1+s_2}{2}} g_{M_2}(s_2)\left(g_{M_1}(s_1 - i\pi) - g_{M_1}(s_1 + i\pi)\right)\Phi(s_1, s_2). \tag{3.26}$$

In the $s_1$ complex plane, the contour of integration can be freely moved between $\text{Im}(s_1) = -\pi$ and $\text{Im}(s_1) = \pi$ because there are no poles and the integral is convergent. Therefore we pull the contour of the $g_{M_1}(s_1 - i\pi)$ part to integrate $s_1$ from $-\infty + i\pi/2$ to $\infty + i\pi/2$, and the contour of the $g_{M_1}(s_1 + i\pi)$ part to integrate $s_1$ from $-\infty - i\pi/2$ to $\infty - i\pi/2$. Then changing the variable $s_1$ to $s_1 - i\pi$ in the $g_{M_1}(s_1 - i\pi)$ part, and $s_1$ to $s_1 + i\pi$ in the $g_{M_1}(s_1 + i\pi)$ part we obtain

$$O_{12} = \frac{i}{4\pi}\left(\int_{-\infty+i\pi/2}^{\infty+i\pi/2} + \int_{-\infty-i\pi/2}^{\infty-i\pi/2}\right)ds_1 \int_{-\infty}^\infty ds_2 \frac{1}{\sinh^2\frac{s_1+s_2}{2}} g_{M_1}(s_1)g_{M_2}(s_2)\psi(s_1, s_2), \tag{3.27}$$

where we have introduced

$$\psi(s_1,s_2) = \int_1^\infty dr\, e^{-\lambda^2\left(\cosh\frac{s_2}{2}\right)^2 r^2} \frac{\cosh\frac{s_1}{2}}{r\sqrt{r^2 - \cosh^2\frac{s_1}{2}}}, \qquad (3.28)$$

which verifies $\Phi(s_1 + i\pi, s_2) = i\psi(s_1,s_2)$ and $\Phi(s_1 - i\pi, s_2) = -i\psi(s_1,s_2)$. We change the variable $r$ to $u$ with $r = \cosh\frac{s_1}{2}/\cosh\frac{u}{2}$ in the integral defining $\psi$, and cut it into two pieces. Using Appendix B for the second piece, we get

$$\int_{\cosh\frac{s_1}{2}}^\infty dr\, e^{-\lambda^2\left(\cosh\frac{s_2}{2}\right)^2 r^2} \frac{\cosh\frac{s_1}{2}}{r\sqrt{r^2 - \cosh^2\frac{s_1}{2}}} = \frac{\pi}{2}\mathrm{Erfc}\left(\lambda\cosh\frac{s_1}{2}\cosh\frac{s_2}{2}\right) \qquad (3.29)$$

the first piece (the remaining integral from 1 to $\cosh\frac{s_1}{2}$) is equal to $+\frac{i}{2}\varphi(s_1,s_2)$ if $\mathrm{Im}(s_1) = +\pi/2$ and $-\frac{i}{2}\varphi(s_1,s_2)$ if $\mathrm{Im}(s_1) = -\pi/2$, with

$$\varphi(s_1,s_2) \equiv \int_0^{s_1} du\, \exp\left(-\lambda^2 \frac{\left(\cosh\frac{s_1}{2}\right)^2\left(\cosh\frac{s_2}{2}\right)^2}{\cosh^2\frac{u}{2}}\right) \qquad (3.30)$$

The Erfc term is exactly the opposite of the even-term given by Equation (3.17). The remaining integral gives a term

$$T_{12} = \frac{i}{8\pi}\left(\int_{-\infty+i\pi/2}^{\infty+i\pi/2} ds_1 - \int_{-\infty-i\pi/2}^{\infty-i\pi/2} ds_1\right)\int_{-\infty}^{\infty} ds_2 \frac{1}{\sinh^2\frac{s_1+s_2}{2}} g_{M_1}(s_1)g_{M_2}(s_2)\varphi(s_1,s_2) \qquad (3.31)$$

and the integral over $s_1$ (provided we close it at infinity) is an integral over a closed loop performed clockwise. The only pole of the integrand with an imaginary part between $-i\pi/2$ and $i\pi/2$ is $s_1 = -s_2$, coming from the term $1/\sinh^2\frac{s_1+s_2}{2}$. Since this loop encompasses only this pole, the value of the integral over $s_1$ is equal to $-2i\pi$ times the residue of the integrand taken at this pole. We will now evaluate it.

Let us note $\partial_i \varphi$ the derivative of $\varphi$ with respect to the i-th variable. Since the pole $s_1 = -s_2$ is of order 2 the residue is

$$\frac{\partial}{\partial s_1}\left(\frac{(s_1+s_2)^2}{\sinh^2\frac{s_1+s_2}{2}} g_{M_1}(s_1)\varphi(s_1,s_2)\right)\bigg|_{s_1=-s_2} = 4\frac{\partial}{\partial s_1}\left(g_{M_1}(s_1)\varphi(s_1,s_2)\right)\bigg|_{s_1=-s_2}$$

$$= 4\left(g'_{M_1}(-s_2)\varphi(-s_2,s_2) + g_{M_1}(-s_2)\partial_1\varphi(-s_2,s_2)\right) \qquad (3.32)$$

and therefore

$$T_{12} = \int_{-\infty}^{\infty} ds_2 g_{M_2}(s_2)\left(g'_{M_1}(-s_2)\varphi(-s_2,s_2) + g_{M_1}(-s_2)\partial_1\varphi(-s_2,s_2)\right)$$
$$= \int_{-\infty}^{\infty} ds_1 ds_2 \delta(s_1+s_2) g_{M_2}(s_2)\left(g'_{M_1}(s_1)\varphi(s_1,s_2) + g_{M_1}(s_1)\partial_1\varphi(s_1,s_2)\right). \quad (3.33)$$

We have to calculate the sum $T = T_{12} + T_{21}$. It is the sum of 2 terms: $T = A + B$, where

$$A \equiv \int_{-\infty}^{\infty} ds_1 ds_2 \left(g'_{M_1}(s_1)g_{M_2}(s_2)\varphi(s_1,s_2) + g_{M_1}(s_1)g'_{M_2}(s_2)\varphi(s_2,s_1)\right)\delta(s_1+s_2)$$
$$= \int_{-\infty}^{\infty} ds_1 ds_2 \left(g'_{M_1}(s_1)g_{M_2}(s_2) - g_{M_1}(s_1)g'_{M_2}(s_2)\right)\delta(s_1+s_2)\varphi(s_1,s_2). \quad (3.34)$$

We have used the fact that when $s_1 + s_2 = 0$, $\varphi$ is transformed into its opposite when $s_1$ and $s_2$ are exchanged. Integrating by parts gives

$$A = -\int_{-\infty}^{\infty} ds_1 ds_2 g_{M_1}(s_1)g_{M_2}(s_2)\left(\frac{\partial}{\partial s_1} - \frac{\partial}{\partial s_2}\right)\{\delta(s_1+s_2)\varphi(s_1,s_2)\}$$
$$= -\int_{-\infty}^{\infty} ds_1 ds_2 \delta(s_1+s_2) g_{M_1}(s_1)g_{M_2}(s_2)(\partial_1\varphi(s_1,s_2) - \partial_2\varphi(s_1,s_2)). \quad (3.35)$$

The second term in $T$ is

$$B \equiv \int_{-\infty}^{\infty} ds_1 ds_2 \delta(s_1+s_2) g_{M_1}(s_1)g_{M_2}(s_2)(\partial_1\varphi(s_1,s_2) + \partial_1\varphi(s_2,s_1))$$
$$= 2\int_{-\infty}^{\infty} ds_1 ds_2 \delta(s_1+s_2) g_{M_1}(s_1)g_{M_2}(s_2)\partial_1\varphi(s_1,s_2) \quad (3.36)$$

(because $\partial_1\varphi$ is invariant by exchange of its variables when $s_1 + s_2 = 0$). Therefore $A + B$ is

$$T = \int_{-\infty}^{\infty} ds_1 ds_2 \delta(s_1+s_2) g_{M_1}(s_1)g_{M_2}(s_2)(\partial_1\varphi(s_1,s_2) + \partial_2\varphi(s_1,s_2)). \quad (3.37)$$

It is straightforward to compute the sum of the 2 partial derivatives of $\varphi$ taken at $s_1 = -s_2$. It yields $\exp(-\lambda^2 \cosh^2(s_2/2))$, and finally

$$T = \int_{-\infty}^{\infty} ds_1 ds_2 \delta(s_1 + s_2) g_{M_1}(s_1) g_{M_2}(s_2) \exp\left(-\lambda^2 \cosh^2 \frac{s_2}{2}\right). \tag{3.38}$$

*c. Calculation of the shrunk-loop term S*

In the shrunk-loop term, the 2 paths have a length $r_1 = r_2 = 2a$, therefore

$$S = \int_{-\infty}^{\infty} ds_1 ds_2 \delta(s_1 + s_2) \exp\left(-\frac{2a^2}{t}(1+e^{s_1})(1+e^{-s_1})\right) g_{M_1}(s_1) g_{M_2}(s_2)$$

$$= \int_{-\infty}^{\infty} ds_1 ds_2 \delta(s_1 + s_2) g_{M_1}(s_1) g_{M_2}(s_2) \exp\left(-\lambda^2 \cosh^2 \frac{s_1}{2}\right) \tag{3.39}$$

and therefore $S = T = O + E$.

### 3.2. 2n scatter case

We follow the case $n = 1$. The shrunk-loop theorem states in the general case that

$$\sum_{\substack{\text{cyclic} \\ \text{perm.}}} E_{12...2n} + \sum_{\substack{\text{cyclic} \\ \text{perm.}}} O_{12...2n} = S_{[M_1 M_2 ... M_{2n}]} \tag{3.40}$$

where $S_{[M_1 M_2 ... M_{2n}]}$ is given by (2.8),

$$E_{12...2n} = \int_{-\infty}^{\infty} \frac{dxdy}{2\pi t} \int_{-\infty}^{\infty} ds_1 ds_2 ... ds_{2n} \exp\left[-\frac{1}{2t}\left(\sum_{k=0}^{2n} r_k e^{s_1+s_2+...+s_k}\right)\left(\sum_{k=0}^{2n} r_k e^{-s_1-s_2-...-s_k}\right)\right]$$

$$\times \frac{1}{(2M_1\pi + i(s_1 + i\theta_1))^2 - \pi^2} \prod_{k=2}^{2n-1} \frac{1}{(2M_k\pi + is_k)^2 - \pi^2} \frac{1}{(2M_{2n}\pi + i(s_{2n} + i\theta_2))^2 - \pi^2} \tag{3.41}$$

and

$$O_{12...2n} = \int_0^{\infty} \int_0^{2\pi} \frac{rdrd\theta}{2\pi t} \int_{-\infty}^{\infty} ds_1 ds_2 ... ds_{2n} ds_{2n+1} \exp\left[-\frac{1}{2t}\left(\sum_{k=0}^{2n+1} r_k e^{s_1+s_2+...+s_k}\right)\left(\sum_{k=0}^{2n+1} r_k e^{-s_1-s_2-...-s_k}\right)\right]$$

$$\times \sum_N \frac{1}{(2N\pi + i(s_1 + i\theta))^2 - \pi^2} \prod_{k=2}^{2n} \frac{1}{(2M_k\pi + is_k)^2 - \pi^2} \frac{1}{(2(M_1 - N)\pi + i(s_{2n+1} - i\theta))^2 - \pi^2}, \tag{3.42}$$

the symbols having the same meaning than in the two-scatter case (and in the even-term, $r_1 = ... = r_{2n-1} = 2a$, and in the odd-term, $r_1 = ... = r_{2n} = 2a$). The summation in (3.40) is over all cyclic permutations of the indices (1,2,...,2n). The proof will follow the same steps as in the two-scatter case. We set $E = \sum\limits_{\text{cyclic perm.}} E_{12...2n}$ and $O = \sum\limits_{\text{cyclic perm.}} O_{12...2n}$.

*a. Reduction of the even-term E*

Starting from Equation (3.41), if we change variables $s_1 + i\theta_1$ to $s_1$ and $s_{2n} + i\theta_2$ to $s_{2n}$, we get

$$E_{12...2n} = \int_{-\infty}^{\infty} \frac{dxdy}{2\pi t} \int_{-\infty+i\theta_1}^{\infty+i\theta_1} ds_1 \int_{-\infty}^{\infty} ds_2...ds_{2n-1} \int_{-\infty+i\theta_2}^{\infty+i\theta_2} ds_{2n} \exp\left(-\frac{R^2}{2t}\right) \prod_{k=1}^{2n} g_k(s_k), \qquad (3.43)$$

where $g_k(s_k) \equiv g_{M_k}(s_k)$ is given by equation (3.6) and $R^2$ can now be written

$$R^2 = -4\left(x\sinh\tau_{2n} - a\sum_{k=1}^{2n}\sinh\tau_k\right)^2 - 4\left(y\sinh\tau_{2n} + ia\sum_{k=1}^{2n}\cosh\tau_k\right)^2. \qquad (3.44)$$

We have introduced the quantities

$$\tau_k = \frac{(s_1 + s_1 + ... + s_k) - (s_{k+1} + ... + s_{2n})}{2}, \quad 0 \le k \le 2n. \qquad (3.45)$$

We can move the contour of the integrals over $s_1$ and $s_{2n}$, and integrate over $x$ and $y$, following the case $n=1$. Introducing $\mu \equiv a\sqrt{2/t} = \lambda/2$, we obtain

$$\frac{1}{2n}\sum_{\text{cyclic perm.}} E_{12...2n}^{y>0} = -\frac{1}{16}\left(\int_{-\infty}^{\infty} ds_1 \int_{-\infty+i\pi/2}^{\infty+i\pi/2} ds_{2n} + \int_{-\infty+i\pi/2}^{\infty+i\pi/2} ds_1 \int_{-\infty}^{\infty} ds_{2n}\right) \int_{-\infty}^{\infty} ds_2...ds_{2n-1} \frac{1}{\sinh^2 \tau_{2n}}$$

$$\times \prod_{k=1}^{2n} g_k(s_k) \text{Erfc}\left(\mu\sum_{k=1}^{2n}\cosh\tau_k\right) \qquad (3.46)$$

and the same formula for $y < 0$ with $+i\pi/2$ replaced by $-i\pi/2$ in the limits of the integral.

*b. Reduction of the odd-term O*

After having done the change of variables:

$$\begin{cases} s_1 + s_{2n+1} \to s_1 \\ s_1 - s_{2n+1} \to s_{2n+1}, \end{cases} \qquad (3.47)$$

the integration over the angle $\theta$ can be performed exactly the same way as in the equations leading to Equation (3.21). We get

$$O_{12...2n} = \frac{i}{4\pi t} \int_0^\infty dr\, r \int_{-\infty}^\infty ds_2...ds_{2n} \int_{-s_1}^{s_1} ds_{2n+1} \exp\left(-\frac{R^2}{2t}\right) \prod_{k=2}^{2n} g_k(s_k) \left[g_1(s_1 - i\pi) - g_1(s_1 + i\pi)\right] \quad (3.48)$$

with

$$R^2 = R^2\left(\frac{s_1 + s_{2n+1}}{2}, s_2, s_3, ..., s_{2n}, \frac{s_1 - s_{2n+1}}{2}\right)$$

$$= 4r^2 \cosh^2 \tau_{2n} + 8ar\cosh \tau_{2n} \sum_{k=1}^{2n} \cosh\left(\tau_k - \frac{s_1 - s_{2n+1}}{2}\right) + 4a^2 \left(\sum_{k=1}^{2n} e^{\tau_k}\right)\left(\sum_{k=1}^{2n} e^{-\tau_k}\right) \quad (3.49)$$

It is convenient to introduce the following quantities:

$$\Omega^2 \equiv \left(\sum_{k=1}^{2n} e^{\tau_k}\right)\left(\sum_{k=1}^{2n} e^{-\tau_k}\right) = \left(\sum_{k=1}^{2n} \cosh \tau_k\right)^2 - \left(\sum_{k=1}^{2n} \sinh \tau_k\right)^2 \quad (3.50)$$

and

$$\begin{cases} \cosh \dfrac{\xi(s_1, s_2, ..., s_{2n})}{2} = \dfrac{1}{\Omega} \displaystyle\sum_{k=1}^{2n} \cosh \tau_k \\ \sinh \dfrac{\xi(s_1, s_2, ..., s_{2n})}{2} = \dfrac{1}{\Omega} \displaystyle\sum_{k=1}^{2n} \sinh \tau_k \end{cases} \quad (3.51)$$

Using Appendix A2, we can change the double integral over $r$ and $s_{2n+1}$ to a single integral. Setting $\mu \equiv a\sqrt{2/t}$, we have

$$\int_{-s_1}^{s_1} ds_{2n+1} \int_0^\infty dr\, r e^{-\frac{R^2}{2t}} = \frac{t}{2\cosh^2 \tau_{2n}} \int_{-s_1}^{s_1} ds_{2n+1} \int_0^\infty dr\, r \exp\left[-\left(\mu^2 \Omega^2 + 2\mu r \sum_{k=1}^{2n} \cosh\left(\tau_k - \frac{s_1 - s_{2n+1}}{2}\right) + r^2\right)\right]$$

$$= \frac{t}{2\cosh^2 \tau_{2n}} \left(\Phi(s_1, s_2, ..., s_{2n}) - \Phi(-s_1, s_2, ..., s_{2n})\right) \quad (3.52)$$

where

$$\Phi(s_1, s_2, ..., s_{2n}) \equiv \int_1^\infty dz\, e^{-\mu^2 \Omega^2 z^2} \frac{\sinh \dfrac{\xi(s_1, s_2, ..., s_{2n})}{2}}{z\sqrt{z^2 + \sinh^2 \dfrac{\xi(s_1, s_2, ..., s_{2n})}{2}}}. \quad (3.53)$$

The odd-term then reads

$$O_{12...2n} = \frac{i}{8\pi} \int_{-\infty}^\infty ds_1...ds_{2n} \frac{1}{\cosh^2 \tau_{2n}} \prod_{k=2}^{2n} g_k(s_k)$$

$$\times \left(g_1(s_1 - i\pi) - g_1(s_1 + i\pi)\right)\left(\Phi(s_1, s_2, ..., s_{2n}) - \Phi(-s_1, s_2, ..., s_{2n})\right). \quad (3.54)$$

Then we pull down the contour of the $g_1(s_1 - i\pi)$ part of the odd-term so that $s_1$ is integrated from $-\infty + i\pi/2$ to $\infty + i\pi/2$, and replace $s_1 + i\pi$ by $s_1$; similarly we pull up the contour of the $g_1(s_1 + i\pi)$ part so that $s_1$ is now integrated from $-\infty - i\pi/2$ to $\infty - i\pi/2$, and then replace $s_1 - i\pi$ by $s_1$. If we define

$$\psi(s_1, s_2, ..., s_{2n}) \equiv \int_1^\infty dr\, e^{-\mu^2 \Omega^2 r^2} \frac{\cosh \frac{\xi(s_1, s_2, ..., s_{2n})}{2}}{r\sqrt{r^2 - \cosh^2 \frac{\xi(s_1, s_2, ..., s_{2n})}{2}}}, \tag{3.55}$$

then $\Phi(s_1 + i\pi, s_2, ..., s_{2n}) = i\,\psi(s_1, s_2, ..., s_{2n})$ (because $\Phi$ and $\psi$ are functions of the $\tau_k$, and all the $\tau_k$ contain the term $+s_1/2$). Therefore we can write

$$O_{12...2n} = \frac{1}{8\pi} \left( \int_{-\infty+i\pi/2}^{\infty+i\pi/2} ds_1 + \int_{-\infty-i\pi/2}^{\infty-i\pi/2} ds_1 \right) \int_{-\infty}^{\infty} ds_2 ds_3 ... ds_{2n} \frac{1}{\sinh^2 \tau_{2n}} \prod_{k=1}^{2n} g_k(s_k)$$
$$\times (\psi(s_1, s_2, ..., s_{2n}) + \psi(-s_1, s_2, ..., s_{2n})). \tag{3.56}$$

Changing the variable $r$ to $u$ with $r = \cosh\frac{\xi}{2} / \cosh\frac{u}{2}$ in the integral defining $\psi$ ($\xi$ stands for $\xi(s_1, s_2, ..., s_{2n})$), and then using Appendix B, we get for the part of the integral going from $\cosh\xi/2$ to $\infty$

$$\int_{\cosh\frac{\xi}{2}}^{\infty} dr\, e^{-\mu^2 \Omega^2 r^2} \frac{\cosh\frac{\xi}{2}}{r\sqrt{r^2 - \cosh^2\frac{\xi}{2}}} = \frac{\pi}{2} \text{Erfc}(\mu\Omega\xi) = \frac{\pi}{2} \text{Erfc}\left(\mu \sum_{k=1}^{2n} \cosh\tau_k\right) \tag{3.57}$$

the remaining integral from 1 to $\cosh\frac{\xi}{2}$ is equal to $+\frac{i}{2}\varphi(s_1, s_2, ..., s_{2n})$ if $\text{Im}(s_1) = +\pi/2$ and $-\frac{i}{2}\varphi(s_1, s_2, ..., s_{2n})$ if $\text{Im}(s_1) = -\pi/2$, with

$$\varphi(s_1, s_2, ..., s_{2n}) \equiv \int_0^\xi du\, \exp\left(-\mu^2 \frac{\left(\sum_{k=1}^{2n} \cosh\tau_k\right)^2}{\cosh^2\frac{u}{2}}\right) \tag{3.58}$$

The Erfc term exactly cancels the "symmetrized" even-term $\frac{1}{2n}\sum E_{12...2n}$ given by Equation (3.46) provided it is summed over cyclic permutations on all indices. The remaining part is

$$T = \sum_{\substack{\text{cyclic} \\ \text{perm}}} \frac{i}{16\pi} \left( \int_{-\infty+i\pi/2}^{\infty+i\pi/2} ds_1 - \int_{-\infty-i\pi/2}^{\infty-i\pi/2} ds_1 \right) \int_{-\infty}^{\infty} ds_2 \ldots ds_{2n} \frac{1}{\sinh^2 \tau_{2n}} \prod_{k=1}^{2n} g_k(s_k)$$

$$\times \left( \varphi(s_1, s_2, \ldots, s_{2n}) - \varphi(-s_1, s_2, \ldots, s_{2n}) \right). \tag{3.59}$$

The integral over $s_1$ is again a loop that encompasses only one pole corresponding to $\sinh \tau_{2n} = 0$, the pole

$$s_1 = -(s_2 + \ldots + s_{2n}). \tag{3.60}$$

Computing the residue as in Equation (3.32) and replacing the condition (3.60) by an integral over a delta function we get

$$T = \sum_{\substack{\text{cyclic} \\ \text{perm}}} \frac{1}{2} \int_{-\infty}^{\infty} ds_1 \ldots ds_{2n} \delta(s_1 + s_2 + \ldots + s_{2n}) \prod_{k=2}^{2n} g_k(s_k) \tag{3.61}$$

$$\times \left[ g'_1(s_1) \left( \varphi(s_1, s_2, \ldots, s_{2n}) - \varphi(-s_1, s_2, \ldots, s_{2n}) \right) + g_1(s_1) \frac{\partial}{\partial s_1} \left( \varphi(s_1, s_2, \ldots, s_{2n}) - \varphi(-s_1, s_2, \ldots, s_{2n}) \right) \right]$$

which is a sum of two terms: $T = A + B$, where $A$ is given by Equation (3.62) and $B$ by Equation (3.65). Let us calculate both terms. We have

$$A \equiv \sum_{\substack{\text{cyclic} \\ \text{perm.}}} \frac{1}{2} \int_{-\infty}^{\infty} ds_1 \ldots ds_{2n} \delta(s_1 + s_2 + \ldots + s_{2n}) \prod_{k=2}^{2n} g_k(s_k) g'_1(s_1) \left( \varphi(s_1, s_2, \ldots, s_{2n}) - \varphi(-s_1, s_2, \ldots, s_{2n}) \right)$$

$$= \sum_{\substack{\text{cyclic} \\ \text{perm.}}} \frac{1}{2} \int_{-\infty}^{\infty} ds_1 \ldots ds_{2n} \delta(s_1 + s_2 + \ldots + s_{2n}) \prod_{k=2}^{2n} g_k(s_k) g'_1(s_1) \left( \varphi(s_1, s_2, \ldots, s_{2n}) - \varphi(s_2, s_3 \ldots, s_{2n}, s_1) \right)$$

$$\tag{3.62}$$

because when $s_1 + s_2 + \ldots + s_{2n} = 0$ we have $\varphi(s_2, s_3, \ldots, s_{2n}, -s_1) = \varphi(s_1, s_2, \ldots, s_{2n})$. Integrating by parts over $s_1$ (or the corresponding $s_k$ in the other permutations) we get

$$A = -\frac{1}{2} \int_{-\infty}^{\infty} ds_1 \ldots ds_{2n} \prod_{k=1}^{2n} g_k(s_k) \left[ \delta(s_1 + s_2 + \ldots + s_{2n}) \sum_{\substack{\text{cyclic} \\ \text{perm.}}} \frac{\partial}{\partial s_1} \left( \varphi(s_1, s_2, \ldots, s_{2n}) - \varphi(s_2, s_3 \ldots, s_{2n}, s_1) \right) \right.$$

$$\left. + \delta'(s_1 + s_2 + \ldots + s_{2n}) \sum_{\substack{\text{cyclic} \\ \text{perm.}}} \left( \varphi(s_1, s_2, \ldots, s_{2n}) - \varphi(s_2, s_3 \ldots, s_{2n}, s_1) \right) \right]. \tag{3.63}$$

The sum over all cyclic permutations in the second line is obviously equal to 0; gathering the remaining terms differently together we obtain

$$A = -\frac{1}{2}\int_{-\infty}^{\infty} ds_1...ds_{2n}\delta(s_1 + s_2 + ... + s_{2n})\prod_{k=1}^{2n} g_k(s_k) \sum_{\substack{\text{cyclic} \\ \text{perm.}}} \left(\partial_{\text{frst}}\varphi(s_1,s_2,...,s_{2n}) - \partial_{\text{last}}\varphi(s_1,s_2,...,s_{2n})\right).$$

(3.64)

where $\partial_{\text{frst}}\varphi$ and $\partial_{\text{last}}\varphi$ are the derivatives of $\varphi$ with respect to the first and the last variable. The second term in $T$ is

$$B \equiv \frac{1}{2}\sum_{\substack{\text{cyclic} \\ \text{perm.}}} \int_{-\infty}^{\infty} ds_1...ds_{2n}\delta(s_1 + s_2 + ... + s_{2n})\prod_{k=1}^{2n} g_k(s_k)\left(\partial_{\text{frst}}\varphi(s_1,s_2,...,s_{2n}) + \partial_{\text{frst}}\varphi(-s_1,s_2,...,s_{2n})\right). \quad (3.65)$$

Using the fact that when (3.60) holds we have $\partial_{\text{frst}}\varphi(-s_1,s_2,...,s_{2n}) = \partial_{\text{frst}}\varphi(s_2,s_3,...,s_{2n},s_1)$, and rearranging the sums, we get for $T = A + B$

$$T = \frac{1}{2}\int_{-\infty}^{\infty} ds_1...ds_{2n}\delta(s_1 + s_2 + ... + s_{2n})\prod_{k=1}^{2n} g_k(s_k) \sum_{\substack{\text{cyclic} \\ \text{perm.}}} \left(\partial_{\text{frst}}\varphi(s_1,s_2,...,s_{2n}) + \partial_{\text{last}}\varphi(s_1,s_2,...,s_{2n})\right). \quad (3.66)$$

Let us calculate the partial derivatives of $\varphi$ with respect to the first and the last variable when the condition $s_1 + s_2 + ... + s_{2n} = 0$ holds. Since $\varphi$ is given by Equation (3.58), we have

$$\frac{\partial}{\partial s_i}\varphi(s_1,s_2,...,s_{2n}) = \frac{\partial \xi}{\partial s_i}\exp(-\mu^2\Omega^2) + \int_0^{\xi} du\, \frac{\partial}{\partial s_i}\exp\left(-\mu^2\frac{\left(\sum_{k=1}^{2n}\cosh\tau_k\right)^2}{\cosh^2\frac{u}{2}}\right). \quad (3.67)$$

Since the $\tau_k$ all contain a term $+s_1/2$ if the sum goes from 1 to $2n$, and all contain a term $-s_{2n}/2$ if the sum goes from 0 to $2n$-1 (which is the same because $\cosh\tau_0 = \cosh\tau_{2n}$), the derivatives $\partial_{s_1}\varphi$ and $\partial_{s_{2n}}\varphi$ of the exp cancel. Furthermore $\Omega$ does not depend on $s_1$ (see Equation (3.50)). The derivatives of $\xi$ give, identifying the derivatives of both sides of Equation (3.51) and considering that $s_1 + s_2 + ... + s_{2n} = 0$,

$$\frac{\partial \xi}{\partial s_1} = 1 \quad \text{and} \quad \frac{\partial \xi}{\partial s_{2n}} = -1 - \frac{2}{\Omega}\frac{\cosh\frac{\xi}{2}}{\sinh\frac{\xi}{2}}\frac{\partial \Omega}{\partial s_{2n}}. \quad (3.68)$$

The derivative of $\Omega$ with respect to $s_{2n}$ has a simple expression when $s_1 + s_2 + ... + s_{2n} = 0$: $\partial \Omega / \partial s_{2n} = -\sinh(\xi/2)$. Therefore

$$T = \int_{-\infty}^{\infty} ds_1...ds_{2n} \delta(s_1 + s_2 + ... + s_{2n}) \prod_{k=1}^{2n} g_k(s_k) \sum_{\substack{\text{cyclic} \\ \text{perm.}}} \frac{\cosh \frac{\xi}{2}}{\Omega} \exp(-\mu^2 \Omega^2). \tag{3.69}$$

But when $s_1 + s_2 + ... + s_{2n} = 0$, $\Omega$ does not depend on the cyclic permutation of the $s_k$. The final step is to notice that

$$\frac{\sum_{\substack{\text{cyclic} \\ \text{perm.}}} \cosh \frac{\xi}{2}}{\Omega} = \frac{\sum_{\substack{\text{cyclic} \\ \text{perm.}}} \sum_{k=1}^{2n} \cosh \tau_k}{\Omega^2} = \frac{\sum_{\substack{\text{cyclic} \\ \text{perm.}}} \sum_{k=1}^{2n} (e^{\tau_k} + e^{-\tau_k})}{2 \left( \sum_{k=1}^{2n} e^{\tau_k} \right) \left( \sum_{k=1}^{2n} e^{-\tau_k} \right)} = 1. \tag{3.70}$$

The final expression for $E+O$ is therefore

$$E + O = \int_{-\infty}^{\infty} ds_1...ds_{2n} \delta(s_1 + s_2 + ... + s_{2n}) \prod_{k=1}^{2n} g_k(s_k) \exp(-\mu^2 \Omega^2). \tag{3.71}$$

*c. Calculation of the shrunk-term S*

The shrunk-term $S_{[M_1 M_2 ... M_{2n}]}$ has an argument in its exponent equal to $-R^2/2t$, where $R$ is given by Equation (2.7), that is

$$-\frac{1}{2t} \left( \sum_{k=1}^{2n} r_k \exp(s_1 + ... + s_k) \right) \left( \sum_{k=1}^{2n} r_k \exp[-(s_1 + ... + s_k)] \right) = -\frac{2a^2}{t} \left( \sum_{k=1}^{2n} e^{\tau_k} \right) \left( \sum_{k=1}^{2n} e^{-\tau_k} \right) = -\mu^2 \Omega^2 \tag{3.72}$$

and the result follows.

**Acknowledgments**
The funding of the Leverhulme trust for A Thain and subsequently for O Giraud is gratefully acknowledged.

# Appendix A1
This appendix shows that for all real $\alpha$

$$I \equiv \int_{-s_1}^{s_1} ds_3 \int_0^\infty dr\, r e^{-(\alpha^2 + 2\alpha r \cosh \frac{s_3}{2} + r^2)} = 2\int_1^\infty dz\, e^{-\alpha^2 z^2} \frac{\sinh \frac{s_1}{2}}{z\sqrt{z^2 + \sinh^2 \frac{s_1}{2}}}. \tag{A1.1}$$

Setting

$$z = \frac{1}{\alpha}\sqrt{\alpha^2 + 2\alpha r \cosh \frac{s_3}{2} + r^2}, \tag{A1.2}$$

we get

$$\begin{aligned}I &= \alpha \int_{-s_1}^{s_1} ds_3 \int_1^\infty dz\, \frac{z}{\sqrt{z^2 + \sinh^2 \frac{s_3}{2}}} \left(-\alpha \cosh \frac{s_3}{2} + \alpha\sqrt{z^2 + \sinh^2 \frac{s_3}{2}}\right) e^{-\alpha^2 z^2} \\ &= s_1 e^{-\alpha^2} + \frac{1}{2}\int_1^\infty dz(-2\alpha^2 z e^{-\alpha^2 z^2})\int_{-s_1}^{s_1} ds_3 \frac{\cosh \frac{s_3}{2}}{\sqrt{z^2 + \sinh^2 \frac{s_3}{2}}}.\end{aligned} \tag{A1.3}$$

Now integrating the $z$ integral by parts, the boundary term cancels the $s_1 e^{-\alpha^2}$ term, and

$$I = \frac{1}{2}\int_1^\infty dz\, e^{-\alpha^2 z^2} \int_{-s_1}^{s_1} ds_3 \frac{z \cosh \frac{s_3}{2}}{\left(z^2 + \sinh^2 \frac{s_3}{2}\right)^{3/2}} = 2\int_1^\infty dz\, e^{-\alpha^2 z^2} \frac{\sinh \frac{s_1}{2}}{z\sqrt{z^2 + \sinh^2 \frac{s_1}{2}}} \tag{A1.4}$$

and the result follows.

## Appendix A2
Noticing that for any $\varphi$ the quantity

$$\Omega^2 = \left(\sum_{k=1}^{2n} \cosh(\tau_k + \varphi)\right)^2 - \left(\sum_{k=1}^{2n} \sinh(\tau_k + \varphi)\right)^2 \tag{A2.1}$$

does not depend on $\varphi$, we can set

$$\begin{cases} \cosh \frac{\hat{\xi}(s_{2n+1})}{2} = \frac{1}{\Omega}\sum_{k=1}^{2n} \cosh\left(\tau_k - \frac{s_1 - s_{2n+1}}{2}\right) \\ \sinh \frac{\hat{\xi}(s_{2n+1})}{2} = \frac{1}{\Omega}\sum_{k=1}^{2n} \sinh\left(\tau_k - \frac{s_1 - s_{2n+1}}{2}\right) \end{cases} \tag{A2.2}$$

where $\hat{\xi}$ stands for $\hat{\xi}(s_1, s_2, ..., s_{2n}, s_{2n+1})$. This appendix shows that

$$I_n \equiv \int_{-s_1}^{s_1} ds_{2n+1} \int_0^\infty dr\, r e^{-(\mu^2\Omega^2 + 2\mu\Omega r \cosh\frac{\hat{\xi}(s_{2n+1})}{2} + r^2)} = \Phi(s_1, s_2, ..., s_{2n}) - \Phi(-s_1, s_2, ..., s_{2n}) \quad (A2.3)$$

where $\Phi(s_1, s_2, ..., s_{2n})$ is defined by

$$\Phi(s_1, s_2, ..., s_{2n}) \equiv \int_1^\infty dz\, e^{-\mu^2\Omega^2 z^2} \frac{\sinh\frac{\xi(s_1, s_2, ..., s_{2n})}{2}}{z\sqrt{z^2 + \sinh^2\frac{\xi(s_1, s_2, ..., s_{2n})}{2}}}. \quad (A2.4)$$

The change of variables

$$z = \frac{1}{\mu\Omega}\sqrt{\mu^2\Omega^2 + 2\mu\Omega r \cosh\frac{\hat{\xi}(s_{2n+1})}{2} + r^2} \quad (A2.5)$$

gives

$$I_n = \mu\Omega \int_{-s_1}^{s_1} ds_{2n+1} \int_1^\infty dz \frac{z}{\sqrt{z^2 + \sinh^2\frac{\hat{\xi}(s_{2n+1})}{2}}} \left(-\mu\Omega\cosh\frac{\hat{\xi}(s_{2n+1})}{2} + \mu\Omega\sqrt{z^2 + \sinh^2\frac{\hat{\xi}(s_{2n+1})}{2}}\right) e^{-\mu^2\Omega^2 z^2}$$

$$= s_1 e^{-\mu^2\Omega^2} + \frac{1}{2}\int_1^\infty dz\left(-2\mu^2\Omega^2 z e^{-\mu^2\Omega^2 z^2}\right) \int_{-s_1}^{s_1} ds_{2n+1} \frac{\cosh\frac{\hat{\xi}(s_{2n+1})}{2}}{\sqrt{z^2 + \sinh^2\frac{\hat{\xi}(s_{2n+1})}{2}}}. \quad (A2.6)$$

Now integrating the $r$ integral by parts, the boundary term cancels the $s_1 e^{-\mu^2\Omega^2}$ term, and

$$I_n = \frac{1}{2}\int_1^\infty dz\, e^{-\mu^2\Omega^2 z^2} \int_{-s_1}^{s_1} ds_{2n+1} \frac{z\cosh\frac{\hat{\xi}(s_{2n+1})}{2}}{\left(z^2 + \sinh^2\frac{\hat{\xi}(s_{2n+1})}{2}\right)^{3/2}} = \int_1^\infty dz\, e^{-\mu^2\Omega^2 z^2} \left[\frac{\sinh\frac{\hat{\xi}(s_{2n+1})}{2}}{z\sqrt{z^2 + \sinh^2\frac{\hat{\xi}(s_{2n+1})}{2}}}\right]_{-s_1}^{s_1} \quad (A2.7)$$

and the result follows since $\hat{\xi}(s_1) = \hat{\xi}(s_1, s_2, ..., s_{2n}, s_1) = \xi(s_1, s_2, ..., s_{2n})$ and $\hat{\xi}(-s_1) = \xi(-s_1, s_2, ..., s_{2n})$.

## Appendix B

This appendix shows that for all real $\alpha$

$$\frac{i}{\pi}\int_0^{i\pi} du\,\exp\left(-\frac{\alpha^2}{\cosh^2\frac{u}{2}}\right) = -\mathrm{Erfc}(\alpha). \tag{B1}$$

Let us define

$$J \equiv \int_0^{i\pi} du\,\exp\left(-\frac{\alpha^2}{\cosh^2\frac{u}{2}}\right) = 2i\int_0^{\pi/2} du\,\exp\left(-\frac{\alpha^2}{\cos^2 u}\right). \tag{B2}$$

Changing variables from $u$ to $t = \tan u$, we get

$$J = 2i\int_0^{\infty} dt\,\frac{1}{1+t^2}\exp\left[-\alpha^2(1+t^2)\right]. \tag{B3}$$

Replacing $1/(1+t^2)$ by $\int_0^{\infty} d\tau\,\exp\left[-\tau(1+t^2)\right]$ and performing the integral over $t$ we get

$$J = i\sqrt{\pi}\int_0^{\infty} d\tau\,\frac{\exp\left[-(\tau+\alpha^2)\right]}{\sqrt{\tau+\alpha^2}}, \tag{B4}$$

which gives $I = i\pi\,\mathrm{Erfc}(\alpha)$ after having set $u = \sqrt{\tau+\alpha^2}$.

## References

[1] Berry M V 1980 Exact Aharonov-Bohm wavefunction obtained by applying Dirac's magnetic phase factor *Eur J Phys* **1** 240-244

[2] Bogomolny E B and Pavloff N and Schmit C 2000 Diffractive corrections in the trace formula for polygonal billiards *Phys Rev E* **61** R3689-R3711

[3] Edwards S F 1967 Statistical mechanics with topological constraints I *Proc Phys Soc* **91** 513-519

[4] Feynman R P and Hibbs A R 1965 *Quanum mechanics and Path integrals* (McGraw Hill)

[5] Hannay J H and Thain A 2003 Exact scattering theory for any straight reflectors in two dimensions *J Phys A: Math Gen* **36** 4063-80

[6] Schwarzschild K 1902 Die Beugung und Polarisation des Lichts durch einen Spalt I *Math Ann* **55** 177-247

[7] Sommerfeld A 1896 Mathematische Theorie der Diffraktion *Math Ann* **47** 317-74

[8] Sommerfeld A 1954 *Optics* (*Lectures on theoretical physics*, *vol 4*) eqn 38.27a (New York: Academic)

[9] Stovicek P 1989 The Green function for the two solenoid Aharonov-Bohm effec*t Phys Lett* A **142** 5-10

[10] Vattay G and Wirzba A and Rosenqvist P E 1994 Periodic Orbit theory of diffraction *Phys Rev Lett* **73** 2304-7